\documentclass[twocolumn,prb,tighten,floatfix,showpacs]{revtex4}
\usepackage{graphicx}

\def\<{\langle}
\def\>{\rangle}

\def\be{\begin{equation}}
\def\ee{\end{equation}}

\begin{document}
\preprint{cond-mat} \title{ Topological entanglement entropy in the second Landau level}

\author{B. A. Friedman and G. C. Levine}

\address{Department of Physics, Sam Houston State University, Huntsville TX 77341}

\address{Department of Physics and Astronomy, Hofstra University,
Hempstead, NY 11549}

\date{\today}

\begin{abstract}The entanglement entropy of the incompressible states of a realistic quantum Hall system in the second Landau level are studied by direct diagonalization. The subdominant term to the area law, the topological entanglement entropy, which is believed to carry information about topologic order in the ground state, was extracted for filling factors $\nu = 12/5$ and $\nu = 7/3$. While it is difficult to make strong conclusions about $\nu = 12/5$, the  $\nu = 7/3$ state appears to be very consistent with the topological entanglement entropy for the $k=4$ Read-Rezayi state. The effect of finite thickness corrections to the Coulomb potential used in the direct diagonalization are also systematically studied. 
\end{abstract}

\maketitle

\section{Introduction}
The physical pictures of states in the  first three Landau levels of quantum Hall systems are quite distinct.  In the lowest, $n = 0$ harmonic oscillator state, Landau level, incompressible states are of most importance, typified by the Laughlin $1/3$ state \cite{tsui} and its descendants from either the hierarchy \cite{haldane} or the composite fermion \cite{jain} approaches.  On the other hand, in the third Landau level $(n=2)$, compressible states, stripes, bubbles or possibly liquid crystal states  are predominate \cite{lilly}.  In many ways, the least understood behavior occurs in  the second $(n=1)$ Landau level.  Here, for example, there is an incompressible state at an even denominator, $ \nu =5/2$ \cite{willett,moore}, an incompressible state at $\nu=7/3$ reminiscent of  the $\nu = 1/3$ Laughlin state and an odd denominator state, $\nu=12/5$ \cite{xia,read}, of mysterious origin.  The present paper is an attempt to better understand these filling factors in the second Landau level using direct diagonalization and the topological entanglement entropy.  In a previous paper \cite{fl}, the topological entanglement entropy (the logarithm of the quantum dimension) was used to argue that the essential physics of the state at $\nu=5/2$ is in fact given by the Moore-Read state.  Here a similar analysis is extended to filling factors $\nu=12/5$ and $\nu=7/3$.

The  basic idea is that the topological entanglement entropy, which of course, is quite distinct from the ordinary thermodynamic entropy, is an invariant that characterizes  incompressible states \cite{kitaev,levin}.  (As a invariant that characterizes a certain "isomorphic" class of objects, the topological entanglement entropy bears a certain resemblance to the Kolmogorov-Sinai entropy of dynamical systems theory. \cite{walters})  Hence, if the value of the invariant for a numerically determined realistic wave function disagrees with a model wave function, that precludes the model wave function.  This approach is complementary to the calculation of the overlap using direct diagonalization, in that agreement with  a "correct" model wave function (or field theory approach), in principle, improves with system size for the topological entanglement entropy.  However, a direct calculation of the overlap gets worse for larger systems, even for the "correct" model wave function \cite{li}.

The paper is organized as follows: in the following section a few numerical issues are discussed, in particular, the role of  finite thickness and how to deal with ground state degeneracy.  In the third section, the results for $\nu=12/5$ and  $\nu=7/3$ are presented, while conclusions are given in the final section.

\section{Numerical Considerations}

The numerical method, developed in ref. [14], [15], and notation used in ref. [ 9] are used in the present study.  The numerical model consists of finite square clusters with periodic boundary conditions with the electrons interacting with long range Coulomb interaction unless otherwise noted.  The electrons are assumed to be fully spin polarized.  Direct diagonalization, implemented by the Lanczos algorithm\cite{dagotto},  is used to find the ground state of the system.  A brief summary of the numerical procedure to find the topological entanglement entropy  is as follows (for full details see ref. [9]).  Firstly, for a given cluster of  $N$ orbitals, the $l$-orbital entanglement entropies  are calculated where $l$ is the number  of orbitals in the subsystem.  The $l$-orbital entanglement entropy is then linearly extrapolated in $1/N$ to large  system sizes; this will be termed, "finite size scaling" (FSS).  The extrapolated $l$-orbital  entanglement entropies are  plotted as a function of $\sqrt{l}$ and fit by a line with the $y$-intercept being the negative of the topological entanglement entropy.   

In finding the ground state, the conserved  quantities  are the total number of electrons and the momentum in the y-direction.  Since we are not doing a full symmetry analysis, the observables of the ground state wave function are not translationally invariant, even though periodic boundary conditions are used.  This leads to a dependence of the $l$-orbital entanglement entropy on the location of the subsystem.  Among the degenerate states with different $k_{y}$'s a state with equal "left" and "right" $l$-orbital entanglement entropies (see  ref. [ 9]) was chosen for which the entanglement entropy was calculated.  This prescription also works, yields a unique state, at $\nu=12/5$ and $\nu=7/3$ and this is the choice we make in section 3.  However, in this section, a different choice will be investigated for $\nu=5/2$, for which  the previous  "recipe" does not always yield a unique state.  The choice made here to guarantee translational invariance of the $l$-orbital entanglement entropy, for a given cluster size, is to merely average the $l$-orbital entanglement entropies for the ground states with different   $k_{y}$'s.  Figure 1 is a graph  of the extrapolated entanglement entropy vs. $\sqrt{l}$ obtained using this procedure.  The diamonds were obtained by using the pure Coulomb potential and the finite size extrapolation (FSS) was done using systems sizes of $N_e = 12,13,14,15$ and $16$ electrons corresponding to $N=24,26,28,30$ and $32$ orbitals.  Note that the electrons in the n=0 Landau level(a spin up and spin down level corresponding to the $2$ in filling factor $\nu=5/2 =2 +1/2$) are treated as inert; only the electrons in the $n=1$ Landau level are considered.  The topological entanglement entropy, $\gamma$, obtained from the linear fit was $\gamma = 2.07 \pm 0.15$.  The standard deviation in the individual extrapolated points, indicated by the error bars, was always less than $0.14$.  This is to be compared to our previous results, where no averaging was done, of $\gamma = 2.01\pm 0.19$ and the maximum error in the extrapolated points being $0.18$.  By either method, there is very good agreement with the topological entanglement entropy of  the Moore-Read  state , i.e. $\gamma = 2.08$.  
\begin{figure}[ht]
\includegraphics[width=6.5cm]{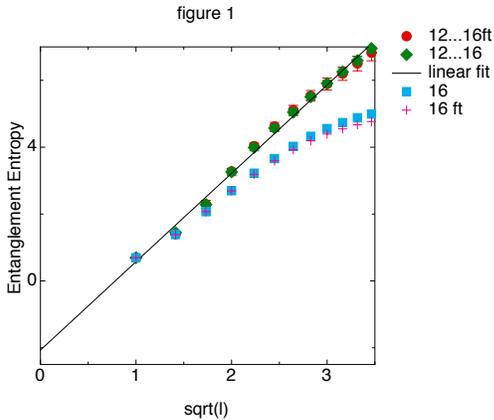}
\caption{\label{fig1} Entanglement entropy of $\nu = 5/2$ state. Entanglement entropy is plotted versus $\sqrt{l}$ (where $l$ is number of orbitals comprising the sub system) for different numbers of orbitals. Squares depict results for the system size with $N_e=16$. Crosses depict the same computation, except with the more realistic finite thickness Coulomb potential.  Finite size scaling (FSS), $N \rightarrow \infty$, results for entanglement entropy versus $\sqrt{l}$ (diamonds) is shown with linear least squares fit (LSF), yielding an intercept of  $-\gamma = -2.07 \pm 0.15$. Circles depict latter computation but using finite thickness Coulomb potential.}
\end{figure}

In figure 1, the squares are the  $l$-orbital  entanglement entropies for the 16 electron system.  Clearly, some sort of system size extrapolation is necessary to obtain a value of the topological entanglement entropy close to  the Moore-Read value.  An additional feature in the above data (the diamonds) is that apart from a linear increase with  $\sqrt{l}$ (the area law) there is a superimposed small sinusoidal like variation.  This is also evident in our previous work on $\nu=1/3$ (and other numerical calculations\cite{morris}).  We suggest that this is not an error in the $1/N$ extrapolation; the asymptotic results of ref . [10] and [11] do not preclude an oscillation which does not grow with the number of orbitals.

Another possibly more realistic choice of interaction, which approximately takes the finite thickness of the physical system into account was  investigated in ref. [18].  The Fourier transform of this "square well" potential is given by

\begin{equation}
V_{sq}(k) = \frac{e^2 l_B }{\epsilon} \frac{1}{k}\frac{1}{(kd)^2+4\pi^2} \{ 3kd +\frac{8\pi^2}{kd} - \frac{32\pi^4(1-e^{-kd})}{(kd)^2((kd)^2+4\pi^2)}\}
\end{equation}

where $d$ is the width of the  quantum well.  The long range (small $k$)  part of this  potential is the same as the Coulomb potential, the short range (large $k$)  part has been modified (softened) by the finite thickness.  The circles in figure 1  are the extrapolated $l$-orbital entanglement entropies and the crosses are the $l$-orbital entanglement entropies for the 16 electron system  both calculated using this potential with $d=4\l_B$, $\l_B$ being the magnetic length.  This choice of $d$ is close to that which gives the maximum overlap with the Moore-Read state for small finite cluster calculations\cite{peterson}.  From the (FSS) extrapolated entropies, one obtains an topologic entanglement entropy of $\gamma = 2.01 \pm 0.19$.  As indicated by the error bars, the extrapolated values have somewhat larger errors (at most $0.24$) for the square well potential.  
\begin{figure}[ht]
\includegraphics[width=6.5cm]{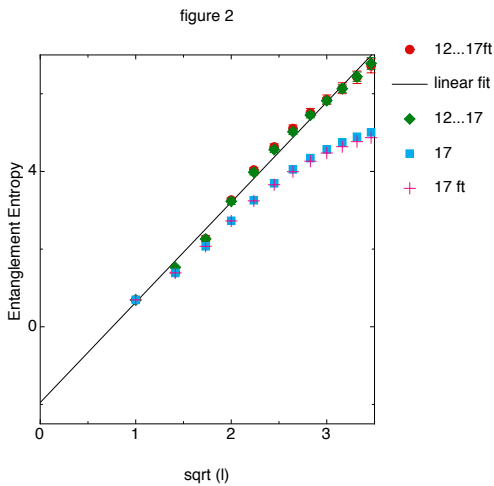}
\caption{\label{fig1} Entanglement entropy of $\nu = 5/2$ state including $N_e=17$ data. Crosses  and squares are entropy vs. $\sqrt{l}$ data for $N_e=17$ with and without finite width Coulomb potential, respectively. Circles and diamonds are the $N_e=12-17$ FSS extrapolated entropies vs. $\sqrt{l}$ data, with and without finite width Coulomb potential, respectively}
\end{figure}

As a further comparison of the square well potential to the pure Coulomb potential, the previous calculation has been extended to $N_e = 17$ electrons in $N = 34$ orbitals, again averaging  over  all ground states.  These calculations involve finding the ground states of matrices of dimension of order $69 \times 10^6$.  The results are shown in figure 2.  The topological entanglement  entropy for pure  Coulomb is $\gamma = 1.93 \pm 0.16$ while the square well gives $\gamma = 1.95 \pm 0.21$.  For either choice of potential, the results agree less closely with the theoretical value of topological entropy for the Moore-Read state.  This is understandable since the Moore-Read state involves pairing of composite fermions which is not as favorable when the number of  electrons  is odd and still relatively small.  The above calculations indicate that as far as the topological entanglement entropy and even the entanglement entropy is concerned, there is not much difference between the square well and ordinary Coulomb interaction.  Assuming that both potentials give the same phase it is not surprising that the topological entanglement entropy is the same for both potentials.  One could hope that the finite size scaling (i.e. the pre asymptotic) behavior of one potential or the other would be better.  However, the square well potential does not seem appreciably better and hence forth we have only considered the pure Coulomb potential.

\section{Topological Entanglement Entropy   of $\nu=12/5$ and $\nu=7/3$}

In this section, the topological entanglement  entropy of the states at  $\nu=12/5$ and $\nu=7/3$ are investigated.  Due to the  possible importance to quantum 
computing\cite{day}, we first consider $\nu=12/5$.  One immediately realizes a problem in that the largest system size that can be easily treated is 14 electrons  in 35 states, involving finding the ground states of matrices of order $66 \times  10^6$.  Thus only 4 system sizes containing 8,10,12, and 14 electrons are included in the extrapolations.  The extrapolated $l$-entanglement entropies, as well as the $l$-entanglement entropies for 14 electrons are  plotted in figure 3.  The topological entanglement entropy, 
\begin{figure}[here]
\includegraphics[width=6.5cm]{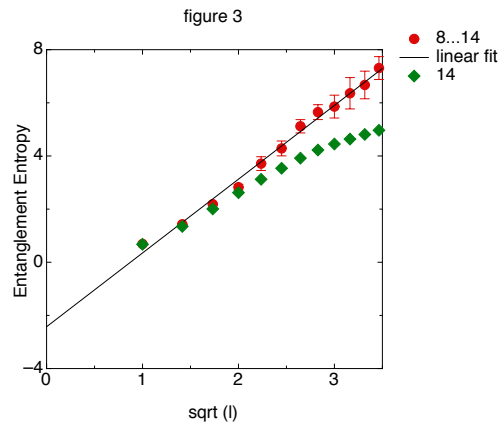}
\caption{\label{fig3} Entanglement entropy of $\nu = 12/5$ state. Diamond symbols depict entropy vs. $\sqrt{l}$ data for $N_e=14$ electrons in $N=35$ orbitals. FSS extrapolation of entropies from $N_e = 8-14$ yields  entropy vs. $\sqrt{l}$ data, depicted with circles. Linear least squares fit (solid line) gives an intercept of  $-\gamma = -2.43 \pm 0.19$ for the $\nu =12/5$ state. }
\end{figure}
obtained from a linear fit, is $\gamma = 2.43 \pm 0.19$.  There are particularly large errors (standard deviations) in the extrapolated values themselves for $l > 8$.  However, this result seems fairly robust in that if one uses only $l$'s up to 8 to get the linear fit,  one still gets a topological entanglement entropy of $2.43$ with a larger standard deviation of  $0.31$.  Taking into account the {\sl two} boundaries for our system on the torus, the $k=3$ Read-Rezayi state has a topological entanglement entropy of $2\ln{D} \approx 2.90$ where $D=\frac{5}{2\sin{\frac{\pi}{5}}}$.  Hence, for our value of $\gamma \approx 2.43$ to be consistent with $k=3$ Read-Rezayi one needs to take the  large possible errors in the extrapolated values into account.  Recall that this  was not necessary at $\nu=5/2$ with the Moore-Read state and the extrapolated errors in the $l$-entanglement entropies  were smaller.  

From an experimental standpoint, the state at $\nu =7/3$ is more robust then that at  $\nu=12/5$ \cite{xia}.  Hence  there is some hope that numerically it will be easier to study the state at $\nu=7/3$.  In figure 4, we plot, as the circles the  extrapolated $l$-entanglement entropy vs. $\sqrt{l}$ for an extrapolation involving 9,10,11 and 12 electrons; the diamonds are an extrapolation for 10,11 and 12 electrons.  The squares are the l-entanglement entropies for the largest system size, 12 electrons.  A linear fit to  the 9-12 electron extrapolations gives an entanglement entropy of $2.92 \pm 0.24$, with rather large error  bars on the last four extrapolated $l$-entanglement entropies.  By way of comparison, the topological entanglement entropy expected from the Laughlin state at $\nu=1/3$ is approximately $1.09$ and for the $k=4$ Read-Rezayi state  is $2\ln{D} \approx 3.58$ where $D=\frac{6}{2 \sin{\frac{\pi}{6}}}$.  Since the error bars in the 10-12 extrapolations are smaller, it is  perhaps more  meaningful to do the linear fit in this  case; one finds a topological  entanglement  entropy of $\gamma = 3.56 \pm 0.33$.  This is in very good agreement with the value expected  for the  $k=4$  Read-Rezayi state.

\begin{figure}[here]
\includegraphics[width=6.5cm]{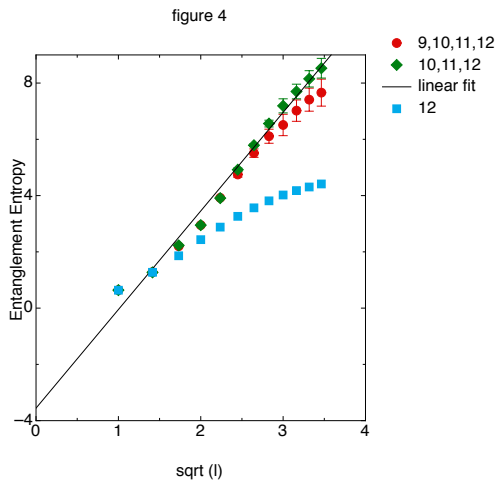}
\caption{\label{fig2} Entanglement entropy of $\nu = 7/3$ state. Same as figure 3 except: squares are $N_e = 12$; diamonds represent FSS extrapolation for $N_e=10-12$; circles represent FSS extrapolation for $N_e=9-12$. Linear least squares fit is performed for $N_e = 10-12$ data (solid line) yielding $\gamma = 3.56 \pm 0.33$ for the $\nu = 7/3$ state. }
\end{figure}
 
\section{Conclusion}
We have calculated the topological entanglement entropies at $\nu=12/5$ and $\nu=7/3$ in the flat torus  geometry.  There are large possible  errors  associated with our results, due to the difficulties  in extrapolating to large system sizes.  That being said, still our calculations tend to support the picture  that $\nu=12/5$ corresponds to the $k=3$ Read-Rezayi state(numerical evidence for this is presented in ref. [20]) and $\nu=7/3$ corresponds to $k=4$ Read-Rezayi (numerical evidence that $\nu=7/3$ is not a Laughlin state is given in ref. [21]) .  Qualitatively, we find that the topological entanglement entropy is  larger for $\nu=7/3$ then for  $\nu=12/5$, this is consistent with assigning these fillings to the $k=4$ and  $k=3$ Read -Rezayi states.  The quantitative case for the Read-Rezayi states  is  stronger for $\nu=7/3$ in that one can get agreement without  taking the  error bars of the extrapolation into account.  Taken  at face value, there is excellent agreement between  our numerical result for $\nu=7/3$, topological entanglement entropy of $3.56 \pm 0.31$ and the value for the $k=4$ Read-Rezayi state $\approx$ 3.58.   Other theoretical alternatives, the Laughlin state or the hierarchical construction (based on $\nu=5/2$) of Bonderson and Slingerson\cite{bonderson}  (topological entanglement entropy of $2 \ln{(2 \sqrt{3})} \approx 2.48 $ \cite{unpub}) have lower topological entanglement entropies which are not consistent with our numerical results.  The $\nu=12/5$ of Bonderson and Slingerland has topological entanglement entropy of $2 \ln (2\sqrt(5)) \approx 3.00 $\cite{unpub}.  This is consistent with our numerical results ( or rather as consistent as the k=3 Read-Rezayi state).  Qualitatively however, this disagrees with the numerics in that the topological entanglement entropies of the Bonderson-Slingerson states increase in going from $\nu=7/3$ to $\nu=12/5$.

If the $k=4$ Read-Rezayi state captures the physics of $\nu=7/3$ we  are left with the following paradoxical situation.  Experiment indicates that the particle-hole conjugate of  $\nu=7/3$,  i.e. $\nu=8/3$ has a quasi particle charge of $e^*=e/3$\cite{dolev}.  Of course, if particle-hole symmetry is  not badly broken this leads to a quasiparticle charge at $\nu=7/3$ of $e^*=e/3$.  However, the quasiparticle charge of $k=4$ Read-Rezayi is $e/6$.  Due to the possibility of breaking particle-hole symmetry there is not necessarily a conflict with experiment and the lack of a incompressible state at $\nu=13/5$ \cite{xia} indicates such symmetry breaking is present.   In any case, a direct experimental measurement of the quasiparticle charge at $\nu=7/3$ would be very interesting.

We thank Parsa Bonderson and Nick Read for helpful correspondence and R.R. Du  for a helpful conversation.  The work, was supported, in part, by the Research Corporation under grant No. CC6535, the Department of Energy, DE-FG02-08ER64623---Hofstra University Center for Condensed Matter (G. L.); the Texas Advanced Research Program Grant No. 003606-00050-2006 and the NSF grant No. DMR=0705048 (B. F.).

\end{document}